\documentclass{aa} 
\usepackage{graphicx}

\usepackage{txfonts} 
\usepackage{natbib}
\bibpunct{(}{)}{;}{a}{}{,}   % to follow the A&A style 
\begin{document} 

\title{The intensity contrast of solar photospheric faculae \\ and network elements II.}
\subtitle{Evolution over the rising phase of solar cycle 23}

\titlerunning{The contrast of solar photospheric faculae and network elements. II.}

   \author{A. Ortiz \inst{1} 
   \thanks{\emph{Present address:} High Altitude Observatory, National Center
	  for Atmospheric Research, P.O. Box 3000, Boulder, CO 80307-3000, USA, \email{ada@ucar.edu}} 
	  \and V. Domingo 
	  \inst{2}
	  \and B. Sanahuja 
	  \inst{1} }

   \offprints{A. Ortiz}

   \institute{Departament d'Astronomia i Meteorologia, Universitat de Barcelona, Mart\'{\i} i
	      Franqu\`{e}s 1, E-08028 Barcelona, Spain\\ \email{Blai.Sanahuja@ub.edu} 
	      \and
	      Institut de Ci\`{e}ncia dels Materials, Grupo de Astrof\'{\i}sica y Ciencias del Espacio,
	      Universitat de Val\`{e}ncia, \\ E-46071 Val\`{e}ncia, Spain \\
	      \email{Vicente.Domingo-Codonyer@uv.es}	      }

   \date{Received / accepted}

   \abstract{

   We studied the radiative properties of small magnetic elements (active region faculae and the
network) during the rising phase of solar cycle 23 from 1996 to 2001, determining their contrasts as a
function of heliocentric angle, magnetogram signal, and the solar cycle phase. We combined
near-simultaneous full disk images of the line-of-sight magnetic field and photospheric continuum
intensity provided by the MDI instrument on board the SOHO spacecraft. Sorting the magnetogram signal
into different ranges allowed us to distinguish between the contrast of different magnetic structures.
We find that the contrast center-to-limb variation (CLV) of these small magnetic elements is independent
of time with a 10\% precision, when measured during the rising phase of solar cycle 23. A 2-dimensional
empirical expression for the contrast of photospheric features as a function of both the position on the
disk and the averaged magnetic field strength was determined, showing its validity through the studied
time period. A study of the relationship between magnetogram signal and the peak contrasts shows that
the intrinsic contrast (maximum contrast per unit of magnetic flux) of network flux tubes is higher than
that of active region faculae during the solar cycle.

   }

   \maketitle

%________________________________________________________________

\section{Introduction} 
\label{intro}

The solar disk is a panoply of magnetic structures that form a hierarchy with a wide range of sizes,
field strengths, and degrees of compactness \citep[e.g.][]{sz00}. The brightness signature of small
magnetic features is a strong function of their heliocentric angle and their size; sunspots are dark
while small flux tubes are generally bright. At high resolution, faculae consist of conglomerates of
tightly packed unresolved bright points with diameters of about 100 km
\citep{dz73,muller83,berger95,berger04}.

Two and a half decades of space-based monitoring of the total solar irradiance have revealed that it
changes on time-scales ranging from  minutes to the length of the solar cycle \citep{willhud88,fl2004}.
Most of these variations arise from the changing presence and evolution of the aforementioned magnetic
features. Variability on the solar rotation time-scale is associated with the passage of sunspots and
faculae across the solar disk \citep[e.g.,][]{foulean86,chap87,lawchap90,solfli02}. These variations can
be as much as  0.3\%. In addition, and superimposed on these variations, an 11-year cycle with a
peak-to-peak amplitude on the order of 0.1\% is now well established by recent total solar irradiance
observations \citep{chap87,foulean88,fro1994,fl2004}. Nowadays, there is a general trend towards
accepting that surface magnetic features can account for most, if not all, of the solar-cycle irradiance
variations \citep[e.g.,][]{solfli02,kri03,walton03,detoma04}; see however, \citet{kld88} or
\citet{sof04}. In particular, the magnetic network has been pointed to as a large contributor to the
solar-cycle variability. This important role has motivated us to study their radiative properties during
the solar cycle.

%------------------------------------------------------------------------------ 
\begin{table*}
\caption{Characteristics of the selected images from 1996 to 2001. The MDI day (first column), date
(second column), and UT time are given for the magnetograms ($t_{mag}$, third column) and their
corresponding averaged intensities ($t_{int}$, fourth column).} 
\footnotesize
\begin{center} 
\begin{tabular}{llll|llll} 
\hline 
\hline 
MDI day & Date & \multicolumn{1}{c}{$t_{mag}$} & \multicolumn{1}{c}{$t_{int}$} & MDI
day & Date & \multicolumn{1}{c}{$t_{mag}$} & \multicolumn{1}{c}{$t_{int}$} \\ 
& & (UT) & (UT) & & & (UT) & (UT)\\ 
\hline 
1237 & 22/05/96 & 00:00 & 00:32 & 2225 & 04/02/99 & 03:11 & 03:11 \\ 
1239 & 24/05/96 & 00:04 *& 01:18 & 2227 & 06/02/99 & 04:47 & 04:47 \\
1249 & 03/06/96 & 12:48 & 12:44 & 2229 & 08/02/99 & 14:23 & 14:23 \\ 
1296 & 20/07/96 & 22:28 *& 22:28 & 2292 & 12/04/99 & 11:12 & 10:31 \\ 
1299 & 23/07/96 & 01:40 *& 01:40 & 2380 & 09/07/99 & 17:35 & 17:32 \\ 
1313 & 06/08/96 & 16:00 & 16:02 &  2413 & 11/08/99 & 09:35 & 09:35 \\ 
1389 & 21/10/96 & 19:11 & 19:12 &  2514 & 20/11/99 & 04:48 & 04:22 \\ 
1407 & 08/11/96 & 19:12 & 18:56 &  2528 & 04/12/99 & 11:11 & 11:11 \\ 
1416 & 17/11/96 & 11:15 *& 11:14 & 2529 & 05/12/99 & 11:11 & 11:11 \\ 
1420 & 21/11/96 & 01:39 *& 01:27  & 2530 & 06/12/99 & 09:35 & 09:35 \\ 
1502 & 11/02/97 & 06:27 *& 06:26 & 2583 & 28/01/00 & 17:35 & 17:27 \\ 
1509 & 18/02/97 & 14:27 *& 14:26 & 2594 & 08/02/00 & 21:32 & 21:33 \\ 
1529 & 10/03/97 & 17:36 & 17:19  & 2652 & 06/04/00 & 17:35 & 17:12 \\ 
1538 & 19/03/97 & 00:00 & 00:08 & 2684 & 08/05/00 & 20:47 & 20:47 \\ 
1572 & 22/04/97 & 11:16 *& 11:16 & 2693 & 17/05/00 & 20:48 & 21:01 \\ 
1614 & 03/06/97 & 11:16 *& 11:16 & 2750 & 13/07/00 & 22:24 & 22:41 \\ 
1701 & 29/08/97 & 03:15 *& 03:14 & 2751 & 14/07/00 & 00:00 & 00:01 \\ 
1717 & 14/09/97 & 16:00 & 16:02 & 2804 & 05/09/00 & 19:11 & 19:11 \\ 
1754 & 21/10/97 & 06:23 & 06:20 & 2860 & 31/10/00 & 17:35 & 17:43 \\ 
1804 & 10/12/97 & 14:23 & 14:23 & 2890 & 30/11/00 & 21:47 & 21:44 \\ 
1882 & 26/02/98 & 18:37 & 18:34 & 2973 & 21/02/01 & 19:11 & 18:56 \\ 
1889 & 05/03/98 & 01:39 *& 02:02 & 3036 & 25/04/01 & 19:12 & 18:44 \\ 
1944 & 29/04/98 & 12:47 & 12:47 & 3099 & 27/06/01 & 16:00 & 16:06 \\ 
1963 & 18/05/98 & 11:11 & 11:11 & 3101 & 29/06/01 & 20:56 & 20:20 \\ 
1968 & 23/05/98 & 06:23 & 06:23 & 3189 & 25/09/01 & 17:39 *& 18:01 \\ 
1983 & 07/06/98 & 03:15 *& 02:30 & 3231 & 06/11/01 & 17:35 & 17:35 \\ 
2000 & 24/06/98 & 11:11 & 11:11 & 3234 & 09/11/01 & 20:47 & 20:53 \\ 
2130 & 01/11/98 & 11:11 & 11:11 & 3237 & 12/11/01 & 22:26 & 22:33 \\ 
2131 & 02/11/98 & 06:23 & 06:23 & 3239 & 14/11/01 & 04:51 *& 05:30 \\ 
2132 & 03/11/98 & 06:23 & 06:23 & 3252 & 27/11/01 & 20:47 & 20:47 \\ 
\hline 
\end{tabular}
\end{center} 
\label{table1} 
\vspace{-0.5cm}
\begin{list}{}{}
\item[*] indicates 5-min averaged magnetograms
\end{list}
\end{table*}
%------------------------------------------------------------------------------

In \citet{paper1}, hereafter Paper I, we determined the contrast of small photospheric bright features
-- active region faculae and the network -- as a function of both heliocentric angle and magnetogram
signal, and we obtained an empirical function that predicts their contrast. The study was performed
using data from ten days during 1999, so it did not consider any temporal evolution with the solar
cycle. Now we consider one more variable -- time -- in order to analyze the solar cycle evolution of the
contrast of these magnetic elements through the rising phase of cycle 23.

Facular and network contrasts are hard to measure, because they come from small bright points -- often
below the resolution limit -- and with a very low contrast. In addition, their brightness signature is a
function of their heliocentric angle, size, averaged magnetic field, wavelength, and spatial resolution
\citep{solanki93,sol94}. Therefore, it is difficult to compare observations made at different
wavelengths, spatial resolutions, or field strengths. We combined cospatial and cotemporal photospheric
intensity images and longitudinal magnetograms with the aim of accurately identifying magnetic features
by their filling factor and of determining the contrast CLV of the different structures sorting them by
their magnetic field. Such contrast measurements are expected to be useful not only for constraining
models of flux tubes, but also for improving the modelling of solar irradiance. Previous works were
essentially photometric studies that did not distinguish features by their magnetic flux
\citep{lk84,wz87,l88,lawchap88,ste96,walton03,ermo03}; these works may suffer a bias towards brighter
features. Relatively few contrast investigations that include the magnetogram signal can be found in the
literature: for example, \citet{fra71}, \citet{ff84}, \citet{topka92,topka97}, \citet{lawtopjo93}, or
Paper I.

We analyzed the temporal evolution of the contrast of small bright features through the rising phase of
cycle 23 (from 1996 to 2001), and continued working on the $\mu^{2}-B^{3}$ model presented in Paper I.
To carry out this investigation we used data from the MDI instrument on board SOHO \citep{dom95}. It was
necessary to perform a careful analysis of the detector's response during the studied period.

In Sect.~\ref{dataproc} we present the data sets used and the analysis procedures. In
Sect.~\ref{results} we describe the results of these investigations, which are discussed in
Sect.~\ref{discu}. Finally, our conclusions are given in Sect.~\ref{conclu}.

%------------------------------------------------------------------------------

\section{Data and analysis procedure} 
\label{dataproc} 
\subsection{Data sets} 
\label{datasets}

The Solar Oscillations Investigation/Michelson Doppler Imager (SOI/MDI) instrument on board the SOHO
spacecraft produces measurements of different observables in the \ion{Ni}{i} 6768 \mbox{\AA} absorption
line. Narrow-band (94 m\mbox{\AA}) filtergrams are obtained at five tuning positions in the vicinity of
the Ni I line by tuning the Michelson's peak transmission. The observables are computed from
combinations of these five filtergrams. The solar image is projected onto a $1024\times1024$ CCD camera,
and the pixel size is $2\times2\mbox{\arcsec}$. This state-of-the-art instrument is described in detail
by \citet{sche95}. Among the observables provided by MDI we are interested in both the full disk
magnetograms and continuum intensity images. Magnetograms measure the line-of-sight component of the
magnetic field averaged over the resolution element, $\langle|\bf B| \cos\gamma\rangle$, where $\gamma$
is the angle between the magnetic field vector and the line-of-sight. For the sake of simplicity, we
hereafter refer to $\langle|\bf B| \cos\gamma\rangle$ as $B$. The continuum intensity is a combination
of the five mentioned filtergrams.

The MDI magnetograms are usually obtained every 96 min, except when 1-min cadence campaigns (for both
magnetograms and continuum intensity images) are carried out. The analyzed data set consists of nearly
simultaneous full-disk magnetograms and continuum-intensity images, recorded during 60 days (10 images
per year) spread over the rising phase of solar cycle 23, and spanning from the 1996 minimum to 2001,
around the solar maximum. These days were chosen because they contain everything from almost field-free
quiet Sun periods to intense activity complexes; the sample contains magnetic activity spread over
almost all $\mu={\rm cos}\,\theta$ values. A detailed list of the selected data and their
characteristics can be found in Table~\ref{table1}. We analyzed the temporal variation of the noise
level in twelve series of the aforementioned 1-min cadence measurements (60 consecutive magnetograms
measured during one hour, see Sect.~\ref{reduction}).

\subsection{Reduction method and analysis} 
\label{reduction}

We employed averages over 5 consecutive intensity images, taken at a cadence of 1 per minute, to reduce
the noise and the signal of the $p-$mode oscillations in the intensity. Individual intensity images were
first corrected for limb-darkening effects, as described in Paper I; then, each image was rotated to
co-align it with the corresponding magnetogram before the averaging. Care was taken to use intensity
images obtained as close in time to the magnetograms as possible. In all sixty cases but six, the two
types of images were recorded within 30 min of each other, and only in one case was the time difference
higher (74 min), because there were no intensities available closer in time to their corresponding
magnetogram. Twenty-three out of the 60 image pairs were exactly simultaneous. Magnetograms are either 1
or 5-min averages, taken at a cadence of 96 min (see Sect.~\ref{dificultades}). Our final data sets are
pairs of co-aligned averaged magnetograms and averaged photospheric continuum intensity images for each
of the 60 selected days that can be compared pixel by pixel.

We determined the 1-$\sigma$ noise level of the MDI magnetograms and continuum images as a function of
position over the CCD array following the procedure described in Paper I, with some slight differences
due to the fact that the present data sets cover half a solar cycle. A fundamental part of this work was
to perform a careful study of the temporal dependence of the MDI sensitivity and stability. In Paper I
we assumed that the MDI noise level was time independent and had remained unchanged between 1996 and
1999. Now we have derived the noise level of the magnetogram images for each year from 1996 to 2001.

We must calculate the temporal dependence of the instrumental noise itself, avoiding potential biases as
much as possible due to the increasing solar activity; it is then imperative to remove any signature of
solar activity. For this, we have compared subsequent averages derived from the twelve 1-min data sets
of magnetograms spread over the six years. Specifically, we took 10 consecutive 1-min magnetograms from
each data set and rotated them to the center of the time series to compensate for differential rotation.
Then, we averaged two groups of five images each, grouping them into odds and evens. Magnetic activity
over 40 G was removed from both averages. The next step was to subtract one 5-min average from the
other, in order to remove magnetic activity as much as possible. Due to the minimal temporal differences
between consecutive magnetograms, solar activity present in both averages should be approximately the
same except for minimum changes in the magnetic configuration. After this subtraction the remaining
fluctuations should be mainly instrumental fluctuations, i.e., noise. Finally, we applied the
$100\times100$ pixels running box described in Paper I to derive the standard deviation of the MDI
magnetograms as a function of position over the CCD array.

\begin{figure}[t]
\centering
\caption{Mean of the standard deviation surfaces (in Gauss) as a function of time, calculated twice
every year between 1996 and 2001. Scatter is small and hardly any trend is visible.} 
\label{fig1} 
\end{figure}

This process was carried out twice every year from 1996 to 2001 to evaluate the temporal change in the
instrumental standard deviation. The resulting standard deviation surfaces,
$\sigma_{\mathrm{mag}}(x,y)$, present an increase towards the SW limb, as discussed in Paper I.
Figure~\ref{fig1} presents the temporal evolution of the noise level between 1996 and 2001 for the 5-min
averaged magnetograms. The rounded-down average value is 8 G, with a variation of 8\% over the six
years; only a slight trend is visible in this time series, which means that the sensitivity of the
detector remains almost constant with time. We chose the median of the two annual standard deviations as
the noise level for that year, $\sigma^i_{\mathrm{mag}}$ ($i$ stands for any of the six years). Note
that the temporal variation of the instrumental noise could be even smaller because the calculated
standard deviations may still contain some noise of solar origin due to magnetic fields that could have
changed significantly within a few minutes.

\begin{figure}[t]
\centering 
\vspace{1cm} 
\vspace{0.3cm}
\caption{Two examples of MDI magnetograms (top panels), their corresponding intensity images after
limb-darkening removal (middle panels), and the resulting contrast masks (lower panels) for activity
minimum (May 22, 1996, left) and maximum (May 17, 2000, right). A circle indicates the size of the Sun
in each contrast mask.} 
\label{fig2} 
\end{figure}

We also analyzed the time evolution corresponding to the mean and standard deviation of the quiet Sun
intensity, $\langle I_{\mathrm{qs}}\rangle$ and $\sigma_{\mathrm{Iqs}}$ respectively. The subscript
$\mathrm{qs}$ denotes ``quiet Sun". Pixels with an absolute magnetic signal value below 0.5 times
$\sigma^i_{\mathrm{mag}}$ were considered as quiet Sun pixels. We evaluated the stability of the mean
non-active Sun intensity with time for the 60 selected days following a procedure similar to that
applied to magnetograms, finding that the intensity fluctuates around 2\% through the analyzed period,
thus also stable enough.

The surface distribution of solar magnetic features with a bright contribution to irradiance variations
is identified by setting two thresholds to every magnetogram-intensity image pair. The first threshold
looks for magnetic activity of any kind, and the second threshold masks sunspots and pores (see a more
detailed explanation in Paper I). Isolated pixels were removed from the set. Then masks were
constructed, for each selected day, that indicate the surface distribution of bright magnetic activity
present over the solar disk at a given moment, as well as the associated contrast for each pixel
$(x,y)$, $C_{\mathrm{fac}}(x,y)$ (see definition in Eq.~1 of Paper I).

In Fig.~\ref{fig2} we show sample magnetograms (top), photospheric continuum intensities (middle), and
their derived contrast masks (bottom) for three days during the rising phase of solar cycle 23. Only
features lying above the given thresholds were pinpointed as black pixels; sunspots, for example, did
not appear in the masks, but their surrounding faculae were identified. For each pixel selected by these
masks, we obtained contrast, magnetic field strength averaged over the pixel, and position (represented
by the heliocentric angle $\mu=\cos\theta$).

\subsubsection{Magnetogram averages} 
\label{dificultades}

The employed magnetograms are a mixture of 1-min and 5-min averages. According to \citet{liu01}, the
noise level of MDI magnetic measurements is 16 G for 1-min magnetograms and 9 G for 5-min magnetograms,
from which we take the 1.77 factor to convert between 1 and 5-min averages. This value does not agree
with the expected value of 2.23 or $\sqrt{5}$, most probably implying that points below the noise level
are not only instrumental noise, but some residuals of solar origin. Before flight the noise level of
single 1-min MDI magnetograms had been estimated to be 20 G \citep{sche95}. After launch, this value was
reduced to 14 G \citep{hag01}. We also derived this later value when performing the standard deviation
procedure to single 1-min magnetograms, and so does \citet{hag01} by other means. Consequently, we
assigned the derived noise level (8 G) whenever the magnetogram had an integration time of 5 min, and
1.77 times that noise (rounded down, 14 G) whenever the magnetogram represented a 1-min measurement.

%------------------------------------------------------------------------------

\section{Results} 
\label{results}

%------------------------------------------------------------------------------ 
\begin{table*}[t]
\caption{Coefficients $a_{j,i}$ of the multivariate fits corresponding to each of the six years.}
\footnotesize 
\begin{center} 
\begin{tabular}{llllllllll} 
\hline 
\hline 
Year & \multicolumn{1}{c}{$a_{01}$} & \multicolumn{1}{c}{$a_{11}$} & \multicolumn{1}{c}{$a_{21}$} &
\multicolumn{1}{c}{$a_{02}$} & \multicolumn{1}{c}{$a_{12}$} & \multicolumn{1}{c}{$a_{22}$} &
\multicolumn{1}{c}{$a_{03}$} & \multicolumn{1}{c}{$a_{13}$} & \multicolumn{1}{c}{$a_{23}$} \\
 & $(\times10^{-4})$ & $(\times10^{-4})$ & $(\times10^{-4})$ & $(\times10^{-6})$ & $(\times10^{-6})$ &
$(\times10^{-6})$ & $(\times10^{-10})$ & $(\times10^{-10})$ & $(\times10^{-10})$ \\ 
\hline 
1996 & $-1.64$ & 11.62 & $-9.18$ & 1.219 & $-2.721$ & 0.638 & $-24.36$ & 15.01 & 22.91 \\ 
1997 & $-2.04$ & 13.86 & $-11.45$ & 1.933 & $-6.385$ & 4.033 & $-33.97$ & 83.00 & $-43.84$ \\ 
1998 & $-1.61$ & 11.71 & $-9.56$ & 1.440 & $-4.363$ & 2.454 & $-19.58$ & 41.10 & $-16.45$ \\ 
1999 & $-2.18$ & 12.08 & $-9.31$ & 0.646 & $-1.753$ & 0.609 & $-2.22$ & $-6.31$ & 13.12 \\ 
2000 & $-1.81$ & 12.87 & $-10.73$ & 0.751 & $-2.972$ & 1.822 & $-6.64$ & 16.02 & $-5.68$ \\ 
2001 & $-0.37$ & 8.98 & $-8.34$ & 0.538 & $-2.520$ & 1.660 & $-5.84$ & 15.38 & $-6.45$ \\ 
\hline 
\end{tabular} 
\end{center} 
\label{table2} 
\end{table*}
%------------------------------------------------------------------------------

The contrast analysis performed here is similar to Paper I, except that now we extend it in time. We
binned $B$/$\mu$ values into eight intervals that range from the threshold level, set at
$3\sigma^i_{\mathrm{mag}}$, to 600 \mbox{G}; this threshold level is, on average, 24 \mbox{G} for 5-min
measurements. Thus, we distinguish between the contrast CLV of magnetic features with different filling
factors by sorting the magnetic flux into different bins. The first four intervals are slightly
different than their equivalent in Paper I in order to account for the new threshold levels.

\begin{figure*}
\centering 
\vspace{0.3cm} 
\vspace{1cm}
\caption{Facular and network contrast at solar minimum (1996) as a function of $\mu$ for eight magnetic
field intervals, from network values (top left panel) to strong faculae (lower right). A dashed line
indicates $C_{\mathrm{fac}}=0$. Solid curves represent a second-degree polynomial least-square fit to
the points; $\mu=1$ is the disk center, $\mu=0$ is the limb. The number in the upper left corner counts
the number of pixels in each $B$/$\mu$ interval.} 
\label{fig3} 
\end{figure*}

Figure~\ref{fig3} represents the contrast, $C_{\mathrm{fac}}$, as a function of $\mu$ for every
$B$/$\mu$ interval during a period of solar minimum (1996). For each $B$/$\mu$ interval -- but the last
-- a second degree polynomial least-squares has been fitted to guide the eye. Numbers in the upper left
corner of each plot indicate the amount of pixels that belong to a given $B$/$\mu$-bin.
Figure~\ref{fig4} represents the same kind of dependence for a period of solar maximum (2001). To avoid
overcrowding we binned data points; as the number of detected pixels grows up toward the solar maximum,
the binning increases from sets of 20 points in 1996 to 40 points in 2001.

\begin{figure*} 
\centering 
\vspace{0.3cm} 
\vspace{1cm}
\caption{Same as Fig.~\ref{fig3} for a solar maximum period (2001).} 
\label{fig4} 
\end{figure*}

These figures, specially Fig.~\ref{fig4}, reveal a clear evolution of the behavior of the contrast from
one $B$/$\mu$ interval to another. Network features (top left panel) show a lower contrast, almost
independent of $\mu$. On the other hand, AR faculae (bottom panels) present a very pronounced CLV.
Features with an intermediate magnetic signal show a gradual increase of the contrast towards the limb, 
as well as an increasingly pronounced CLV. Note the shift in the contrast peak towards lower values of
$\mu$ for increasing values of  $B$/$\mu$. Low $B/\mu$ values ($< 200$ \mbox{G}) always report a
positive contrast everywhere, while it becomes negative around disk  center for $B$/$\mu\geq200$
\mbox{G}.

By examining the temporal series of $C_{\mathrm{fac}}$ as a function of $\mu$, it is easy to notice the
increase in solar activity along the rising phase of the cycle (just looking at the number of pixels
involved). While the highest $B$/$\mu$ bins at solar minimum just contain a few points, plots are
overcrowded at solar maximum. For example, low magnetic signals ($B$/$\mu < 90$ \mbox{G}, most probably
belonging to the network), multiply by 2.5 from 1996 to 2001, while high magnetic signals ($B/\mu > 400$
\mbox{G}, corresponding to strong faculae) increase by a factor of 70. This is a direct consequence of
the growing number of active regions present over the solar disk as solar maximum approaches.

%------------------------------------------------------------------------------
\begin{table*}[t]
\caption{Computed $t$-values from the comparison of the contrast for different years with respect to the
activity minimum contrast (1996, otherwise indicated), for the $B/\mu$ bins shown in Figs.~\ref{fig3}
and \ref{fig4}.}
\footnotesize 
\begin{center} 
\begin{tabular}{llllll} 
\hline 
\hline 
$B/\mu$ & 1997 & 1998 & 1999 & 2000 & 2001 \\
\hline 
$<$ 60 G & 0.05 & 0.21 & 1.18 & 0.40 & 0.21 \\
60 - 90 G & 0.07 & 0.18 & 1.24 & 0.52 & 0.11 \\
90 - 130 G & 0.07 & 0.10 & 1.24 & 0.59 & 0.05 \\
130 - 200 G & 0.03 & 0.22 & 0.99 & 0.52 & 0.10 \\
200 - 300 G & 0.43 & 1.02 & 0.02 & 0.19 & 0.77 \\
300 - 400 G$^{\mathrm{*}}$ & - & 1.01 & 0.30 & 0.27 & 0.81 \\
400 - 500 G$^{\mathrm{**}}$ & - & - & 0.12 & 0.20 & 0.30 \\
500 - 600 G$^{\mathrm{**}}$ & - & - & 0.63 & 0.40 & 0.09 \\
\hline 
\end{tabular} 
\begin{list}{}{}
\item[] \hspace{4.6cm} \tiny $^{\rm(*)}$~Comparison with respect to 1997 ; $^{\rm(**)}$ Comparison with
respect to 1998\\
\end{list}
\end{center} 
\label{table3} 
\end{table*}
%------------------------------------------------------------------------------

Following the method detailed in Paper I, we performed a multivariate analysis in order to obtain an
empirical expression for the contrast of photospheric features, both as a function of $\mu$ and
magnetogram signal, $C_{\mathrm{fac}}(\mu,B/\mu)$. In this work we extend the analysis through the
rising phase of solar cycle 23. Again, a $(\mu, B/\mu)$ grid is used, where grid dimensions are
\mbox{$0.1 \leq \mu \leq 1$} and \mbox{$24\,\mbox{G}\leq(B/\mu)\leq630\,\mbox{G}$}. The $\mu$ values
have been binned linearly, with $\Delta\mu=0.1$. The $B$/$\mu$ bins were chosen to be equally spaced on
a logarithmic scale, with $\Delta\log(B/\mu)=0.05$, in order to compensate for the fact that magnetic
signals are mostly concentrated towards lower values (Figs.~\ref{fig3} and \ref{fig4}). We used the
fitting function applied in Paper I (a second-order polynomial function of $\mu$ and a cubic function of
$B/\mu$), with the contrast constrained to go through zero when $B$/$\mu$=0, which is the expected
behavior for the quiet Sun. The terms of the expression for the contrast $C_{\mathrm{fac}}(\mu,B/\mu)$
can be grouped as:

\begin{eqnarray}  
C_{\mathrm{fac}}(\mu, B/\mu) & = &
\left[a_{01}+a_{11}\mu+a_{21}\mu^{2}\right]\left(\frac{B}{\mu}\right)+ \label{eq1} \\ & &
\left[a_{02}+a_{12}\mu+a_{22}\mu^{2}\right]\left(\frac{B}{\mu}\right)^{2}+ \nonumber \\  &
&\left[a_{03}+a_{13}\mu+a_{23}\mu^{2}\right]\left(\frac{B}{\mu}\right)^{3}. \nonumber 
\end{eqnarray}

\noindent where $a_{ji}$ are the coefficients of the fit for each year and are given in
Table~\ref{table2}. The result of these fits for each of the six years are surfaces that represent the
contrast of bright features. The validity of Eq.~(\ref{eq1}) is limited by the MDI parameters, such as
wavelength and spatial resolution, 6768 \mbox{\AA} and $2\mbox{\arcsec}$, respectively. Other values of
these parameters would result in different contrast dependences on $\mu$ and $B$/$\mu$.

\begin{figure*}[ht]
\centering 
\vspace{-0.5cm}   
\vspace{1cm}
\caption{Differences of yearly contrasts respect to the 1996 contrast as a function of $\mu$ (left column)
and as a function of $B$/$\mu$ (right column), for sample magnetic field ranges and positions over the
solar disk. Each curve represents the difference, for each year, between the calculated contrast from
Eq.~(\ref{eq1}) and the 1996 contrast, in order to search for contrast changes relative to the
activity minimum.}    
\label{fig5}   
\end{figure*}

As can be seen in Figs.~\ref{fig3} and \ref{fig4}, it appears that the contrast CLV of small magnetic
features does not change significantly with time. To quantitatively verify this impression, we performed
a Student's-$t$ test to compare the contrasts for different years relative to the contrast during the
activity minimum of 1996. We calculated cuts to the yearly contrast surfaces in both variables, $\mu$
and $B$/$\mu$, and found that the absolute differences between the contrast for 1997, 1998, 1999, 2000,
and 2001 with respect to the minimum are smaller than 0.01, being the relative differences smaller than
10\% (the higher values appear near the limb and are due to the associated foreshortening).
Figure~\ref{fig5} shows two examples of these differences between different years with respect to 1996.
These examples correspond to some $B$/$\mu$ intervals shown in Figs.~\ref{fig3} and \ref{fig4}, and are
representative of the contrast comparison for $C_{\mathrm{fac}}(\mu)$ and $C_{\mathrm{fac}}(B/\mu)$
(left and right panels, respectively). The Student's-$t$ test reveals that, at the 95\% confidence
level, there is no significant difference between contrast means for different years, and the one during
activity minimum. Moreover, these differences are neither arranged in any specific temporal order nor do
they follow any pattern, as can be seen in Fig.~\ref{fig5}. In order to compare these differences among
contrasts, Table~\ref{table3} presents the calculated $t$-values that result from the test, for the same
magnetic flux bins as shown in Figs.~\ref{fig3} and \ref{fig4}. The tabulated $t$-value (for 16 degrees
of freedom) and probability p = 0.05 (95\% confidence level) is 2.12. Therefore we conclude that, with a
10\% precision, there is no temporal trend for the contrast CLV. Since the nature of the observed
contrasts is related to the structure of the underlying flux tubes, this supports the idea that the
physical properties of the facular flux tubes do not vary with time, in particular with the solar cycle.

The dependence of the peak of $C_{\mathrm{fac}}$ on $B$/$\mu$ is shown in Fig.~\ref{fig6}. This figure
is similar to Fig. 8 in Paper I, but this time we consider the temporal evolution of the small
photospheric magnetic elements. Different symbols are representative of different years. For each year,
the $\mu$-values at which $C_{\mathrm{fac}}$ peaks are plotted against the corresponding magnetic signal
in Fig.~\ref{fig6}a. Figure~\ref{fig6}b shows the maximum $C_{\mathrm{fac}}$ values,
$C_{\mathrm{fac}}^{\mathrm{max}}$, plotted as a function of $B$/$\mu$. These maximum contrasts are
derived from the corresponding modeled surfaces. Finally, Fig.~\ref{fig6}c shows
$C_{\mathrm{fac}}^{\mathrm{max}}/(B/\mu)$ represented as a function of $B$/$\mu$, where
$C_{\mathrm{fac}}^{\mathrm{max}}/(B/\mu)$ is the specific contrast (maximum contrast per unit magnetic
flux). Errors in $\mu$ and $C_{\mathrm{fac}}^{\mathrm{max}}$ are estimated from the difference between
the peak of the fitted surfaces and the peak obtained directly from the data points. Error bars plotted
in Fig.~\ref{fig6} represent an average over the annual errors. Figure~\ref{fig6}c proves that the
contrast per unit of magnetic signal decreases with an increasing magnetic signal. Individual flux tubes
cannot be resolved by MDI pixels, therefore we cannot infer their intrinsic contrast from
Fig.~\ref{fig6}b. However, by normalizing by $B$/$\mu$, we obtain a magnitude roughly proportional to
the intrinsic brightness of the elemental flux tubes, assuming that the field strength of the elemental
magnetic flux tubes lies in a narrow range. Figure~\ref{fig6} (especially Fig.~\ref{fig6}b) also
reflects that contrasts do not change significantly along the present solar cycle. Notice that these
results apply to the intensity of the continuum measured at the \ion{Ni}{i} 6768 \mbox{\AA} line with a
pixel size of $2\mbox{\arcsec}$ and a resolution of $4\mbox{\arcsec}$. There are indications
\citep{dom2005} that the maximum contrast is found at lower values of $\mu$ when the measurements are
performed with higher resolution.

A comparison of Figs.~\ref{fig3} and ~\ref{fig4} above and Fig.~3 in Paper I shows that there are some
differences in the CLV of the contrast. The contrasts obtained in Paper I are systematically higher and
show a different contrast CLV behavior for low $B$/$\mu$ fluxes. These differences are due to the fact
that in Paper I a common quiet Sun intensity background was used, while in the present study the
non-active Sun backgrounds are calculated individually for every observed day. To prove it, we
recalculated the contrasts shown in Paper I with the quiet Sun background used there, as well as with
individual quiet Sun backgrounds, obtaining the same observed differences.

%------------------------------------------------------------------------------

\section{Discussion}
\label{discu}
\subsection{Comparison with previous observations}

According to \citet{pap02}, it is not known whether the contrast of the network changes as a function of
wavelength, position on the disk, and phase in the solar cycle. This is precisely the topic addressed in
this work. We have seen that the contrast of small flux tubes -- those producing a bright contribution
to the total irradiance -- depends on their position on the visible disk and on the magnetic flux, but
do not vary with time. 

Comparison with other contrast center-to-limb observations is not easy because of the differences in the
selected wavelength, spatial resolution, range of studied heliocentric angles, magnetic filling factor,
and size of the analyzed features. All these factors contribute to the scatter between the existing
contrast measurements. Our results differ from earlier observations of the contrast of bright features,
especially when considering magnetic signals $B$/$\mu > 200$ G at disk center. Previous measurements of
disk center facular contrasts have frequently yielded positive values, while in Paper I we measured
negative contrasts at disk center for faculae within the range $200 < B$/$\mu < 600$ G, as did, for
example, \citet{topka92,topka97} or \citet{lawtopjo93} at other spatial resolutions and wavelengths. In
this work we show that this is still true at different phases of solar activity. Measuring the contrast
at disk center is particularly important because it helps in determining the thermal structure of the
deep layers.

The variation in the contrast of small magnetic features with the solar cycle has hardly been
investigated. To our knowledge, only \citet{ermo03} and this work address this problem at the
photospheric level. These authors use Rome-PSPT images to analyze the quiet network pattern during the
current solar cycle and report a network contrast change of about 0.05\% during that period of time. A
possible explanation for the discrepancy with the results presented here is that their identification
method differs substantially from ours: they only use photometric intensity images, while we combine
magnetograms with intensity images to search for the desired features. Thus, different selection
criteria yield to different considerations of what the network is. In addition, they restrict their work
to the quiet network pattern, while we consider a broader population of small magnetic elements. In
particular, the contrast change identified by these authors can be due to a change in the distribution
of elements within the network component with the solar cycle or, similarly, to an increase in the
filling factor of their selected pixels from minimum to maximum; as the maximum is approached, more
network tubes would be included within a pixel, increasing the brightness of such pixel. Also, our 10\%
precision could well include the small change that these authors report. At the chromospheric level,
\citet{worden98} also find that the intensity contrasts of the plage, enhanced network, and active
network remain approximately constant over the solar cycle.

It is remarkable that an expression as given by Eq.~(\ref{eq1}) reproduces the dependence of the
contrast of bright features on their position ($\mu$) and on the magnetic flux per pixel ($B$/$\mu$),
within the range \mbox{$0.1 \leq \mu \leq 1$} and \mbox{$24\,\mbox{G}\leq(B/\mu)\leq630\,\mbox{G}$}. A
relative accuracy of better than 12\% is achieved almost everywhere within this domain and in the period
studied. We are aware that more work needs to be done, since other parameters on which the contrast
depends are kept fixed, such as spatial resolution and wavelength. The $\mu^{2}-B^{3}$ description of
the facular and network contrast has proved valid for different phases of the present solar cycle and
degrees of solar activity, and can be introduced as one input into different models that reproduce the
facular irradiance \citep[see, e.g.,][]{lock05}. Ideally, models that reproduce the facular contribution
to the total solar irradiance variations should rely on broadband photometric observations of the
facular contrast. However, almost every facular contrast measurement found in the literature is
monochromatic. \citet{fou04} have recently shown that this problem can be circumvented by applying a
blackbody correction to monochromatic facular contrasts, thus converting such measurements a posteriori
into bolometric contrasts.

\begin{figure}[ht]
\vspace{-1.cm}  
\hspace{-0.2cm} 
\vspace{0.5cm}  
\caption{Dependence on the corrected magnetic flux per pixel, $B/\mu$, of: {\bf a)} $\mu_{\mathrm{max}}$; {\bf b)}
$C_{\mathrm{fac}}^{\mathrm{max}}$ times $10^{2}$; {\bf c)} $C_{\mathrm{fac}}^{\mathrm{max}}/(B/\mu)$
times $10^{4}$. Each symbol is representative of a different year (see legend). Error bars represent an
average over the annual errors; for clarity, they have been shifted down.} 
\label{fig6}
\end{figure}

\subsection{Theoretical considerations}
\label{teo}

We have shown that the intensity contrast of small photospheric magnetic elements is time independent
during the studied period. Since the nature of the observed contrast CLV is directly related to the
structure of the flux tubes making up these magnetic elements, this observed invariance with time
supports the idea that the physical properties of the facular flux tube do not vary with time, in
particular with the solar cycle. This temporal invariability has often been assumed in the past -- there
is no theoretical argument that implies that small magnetic features should have a different structure
at solar minimum than at maximum -- but has never before been verified. Interestingly enough,
\citet{alt84} report that the umbra contrast is a linear function of the phase in the solar cycle,
yielding changes in the contrast of sunspots of around 30\% along the cycle. Such a temporal change
should have been detected with the precision achieved here; however, small magnetic elements seem to
present a different behavior along the solar cycle than sunspots, as seen in this work.

If the contrast of the studied features does not vary with time, a possible explanation to the increase
in the solar irradiance during solar activity maximum \citep[or for the increase in the contrast found
by][]{ermo03} is the large increase in the number of features at solar maximum. The ratio between the
number of magnetic features at maximum relative to that at minimum increases with $B$/$\mu$ or,
alternatively, with the size of the structure. For example, from Figs.~\ref{fig3} and \ref{fig4} we
found that features with $B$/$\mu < 90$ \mbox{G} increased by a factor of 2.5 from 1996 to 2001.
\citet{meu03} finds a factor of 2.9 for fluxes below $3\,10^{19}$ Mx, which approximately correspond to
network patches. High magnetic signals ($B$/$\mu > 400$ \mbox{G}) increased by a factor of 70 from 1996
to 2001, as a direct consequence of the growing number of active regions present over the disk when
reaching the maximum.

The upper panels of Figs.~\ref{fig3} or ~\ref{fig4} refer to, on average, small flux tubes that dominate
the network population, while the lower panels of those figures refer to the larger tubes mostly present
in AR faculae. There are clear differences between small network flux tubes and tubes found in AR
faculae, i.e. regions with larger $B$/$\mu$. Network tubes are bright everywhere on the solar disk and
exhibit a low contrast (Fig.~\ref{fig6}b) but a high specific contrast (Fig.~\ref{fig6}c). Somewhat
larger tubes are predicted to appear dark at disk center but bright near the limb \citep{ks88}; we can
see this behavior in the lower panels of Fig.~\ref{fig4}. Their contrast is higher than for smaller
tubes (Fig.~\ref{fig6}b) but show a lower specific contrast (Fig.~\ref{fig6}c). This behavior for the
specific contrast was already shown in Paper I, but here we extend its validity for the rising phase of
solar cycle 23; and there is no reason to think that this will change through the remaining solar cycle.
This observed behavior implies that network flux tubes are intrinsically brighter than AR flux tubes
along the solar cycle. The greater brightness at disk center implies that network flux tubes may have a
hotter bottom than larger flux tubes, which follows directly from Figs.~\ref{fig3} and \ref{fig4} and
was predicted by \citet{spruit76}. Another difference between network and facular populations is that
the higher the magnetic signal, the lower the $\mu$-value at which the contrast peaks (see
Fig.~\ref{fig6}a), so that network-like features dominate at disk center and features with larger
$B$/$\mu$ closer to the limb. This is true for the studied period.

Finally, the high specific contrast of small $B$/$\mu$ features (Fig.~\ref{fig6}c) during the solar
cycle, and the fact that their contrast is positive over the whole solar disk (i.e. at all $\mu$'s),
makes us think that a change in the distribution of the network -- for example with the solar cycle --
makes a significant contribution to the change of the total solar irradiance \citep[see,
e.g.,][]{fl2004}. Since the contrast CLV is an input for some models that reproduce the facular
contribution to irradiance variations, and we have shown that the $\mu^{2}-B^{3}$ empirical description
for the facular contrast is valid along the solar cycle, this description can in principle be used to
reproduce irradiance variations for any time within the present solar cycle \citep{wenzler02,lock05}, at
least up to 10\% precision.

%------------------------------------------------------------------------------

\section{Conclusions} 
\label{conclu}

We have presented an analysis of the intensity contrast of small bright magnetic elements -- from AR
faculae to the network -- and its dependence on position, magnetic field strength, and phase of the solar
cycle using MDI/SOHO data. These magnetic features can play an important role in the solar cycle
irradiance variations, which are not fully understood.

While in Paper I we showed that the contrast CLV of magnetic features changes gradually with magnetogram
signal, we show here that this behavior for $C_{\mathrm{fac}}(\mu)$ remains during the solar cycle. Our
results indicate that the contrast of small bright features does not present a temporal dependence with
solar cycle phase, with an uncertainty of 10\%. From this result we infer that the physical properties
of the underlying flux tubes do not vary with time.

The validity of the simple expression presented in Eq.~(\ref{eq1}) -- the $\mu^{2}-B^{3}$ model to
predict the contrast of bright features -- has been extended to the rising phase of solar cycle 23.

We have shown that the intrinsic contrast of network flux tubes is higher than that of AR faculae during
the solar cycle; moreover, the network always makes a positive contribution to the irradiance.
Therefore, it is reasonable to conclude that any change in the network distribution may have a
significant impact on the variation of the total irradiance.

Finally, we conclude that a very careful analysis of the response of the detector with time is
imperative in any study of the long-term radiative properties of faculae and the network. It is
essential to accurately determine the quiet Sun intensity backgrounds and noise levels.

Next steps in this investigation would include, but would not be limited to, introducing the
$\mu^{2}-B^{3}$ model (corrected for bolometric measurements) in reproductions of the solar cycle
irradiance variations, and separating the contributions from faculae, network, and other magnetic
structures in order to analyze the relative role of each of them. Of special interest would be to
compare the role of AR faculae and the network in these long-term irradiance variations.

%________________________________________________________________

\begin{acknowledgements}

Part of this work was supported by the Spanish Ministry of Science and Technology through project AYA2001-3304 and AYA2004-03022. AO
acknowledges financial support from the DURSI (Generalitat de Catalunya) grant 2001 TDOC00021. S. Criscuoli is acknowledged for helpful
discussions. We thank the referee for valuable comments that led to the improvement of the manuscript.

\end{acknowledgements}

%________________________________________________________________

\end{document}